\documentclass[preprint,12pt]{elsarticle}

\usepackage{graphics}
\usepackage{amssymb}
\journal{Journal of Crystal Growth}

\begin{document}

\begin{frontmatter}



\title{Vanishing linear term in chemical potential difference in volume term of work of critical nucleus formation for phase transition without volume change}


\author{Atsushi Mori\footnote{Corresponding author: E-mail atsushimori@tokushima-u.ac.jp, Tel +81-88-656-9417, Fax +81-88-656-9435} and Yoshihisa Suzuki}

\address{Institute of Technology and Science, The University of Tokushima,
2-1 Minamijosanjima, Tokushima 770-8506, Japan}

\date{November 24. 2012; revised April 1, 2013}

\begin{abstract}
A question is given on the form $n(\mu_\beta-\mu_\alpha)$ for the volume term of work of formation of critical nucleus.
Here, $n$ is the number of molecule undergone the phase transition, $\mu$ denotes the chemical potential, $\alpha$ and $\beta$ represent the parent and nucleating phases, respectively.
In this paper we concentrate phase transition without volume change.
We have calculated the volume term in terms of the chemical potential difference $\mu_\mathit{re}-\mu_\mathit{eq}$ for this case.
Here, $\mu_\mathit{re}$ is the chemical potential of the reservoir and $\mu_\mathit{eq}$ that at the phase transition.
We have
\[
W_\mathit{vol} = -\frac{\kappa_\beta-\kappa_\alpha}{2v_\mathit{eq}^2} (\mu_\mathit{re}-\mu_\mathit{eq})^2 V_\beta
\]
with $\kappa$ denoting the isothermal compressibility, $v_\mathit{eq}$ being the molecular volume at the phase transition, $V_\beta$ the volume of the nucleus.
\end{abstract}


\end{frontmatter}

\newpage

% \linenumbers

\section{Introduction}
\label{sec:intro}
To calculate the reversible work of formation of the critical nucleus is one of the main purposes of the theory of nucleation, because one can predict the steady-state nucleation rate $J_{s} = J_0 \exp(-W^*/k_BT)$ through the work of formation of the critical nucleus $W^*$, where $k_BT$ is the temperature multiplied by Boltzmann's constant.
Here, $W^* \equiv W(R^*)$ is the height of the work of formation of critical nucleus with $R^*$ being the radius of the critical nucleus.
We often encounter the following formula, Eq.~(\ref{eq:common}), or equivalent one:
\begin{equation}
\label{eq:common}
W = n(\mu_\beta - \mu_\alpha) + \gamma A,
\end{equation}
with $\gamma$ begin the interfacial tension, $A \equiv 4\pi R^2$ the area of the interface (rigorously speaking, $R$ is the radius of the surface of tension) in textbooks such as Refs.~\cite{wunderlich,chernov,mutaftschiev,kashchiev,markov,kelton} as well as research papers such as Refs.~\cite{auer2001,gasser2001,kashchiev2004,merikanto2007,wedekind2008,kashchiev2010,kawasaki2010}.
In this formula, one regards $\mu_\beta - \mu_\alpha$ as the chemical potential difference between the parent phase (the $\alpha$ phase) and the nucleating phase (the $\beta$ phase).
One can understand $n$ as the numbers of molecules undergone the phase transition from the $\alpha$ phase to $\beta$ phase.
It seems to imply that no volume change is assumed to be associated with the $\alpha$-$\beta$ phase transition.
That is, $\Delta(n\mu)$ reduces to $n\Delta\mu$ in the case that $n$ is common to both $\alpha$ and $\beta$ phases; if $n_\alpha \equiv V_\beta/v_\alpha$ and $n_\beta \equiv V_\beta/v_\beta$ with $V_\beta \equiv 4\pi R^3/3$, one has $n_\alpha = n_\beta (\equiv n)$ for the case that the molecular volumes $v_\alpha$ and $v_\beta$ equals with each other.
In crystal growth from the melt, the difference between densities of the crystal and melt phases is often neglected.
Indeed, Eq.~(\ref{eq:common}) has been used for crystal nucleations in the melt.
However, while someone describes Eq.~(\ref{eq:common}) or equivalent one in phase transitions with small volume change such as melt-crystal cases, someone does in vapor-liquid cases.
In some literatures, implication is unrevealed even by reading between lines.

The exact form for $W$ given by Gibbs~\cite{gibbs} is
\begin{equation}
\label{eq:exact}
W = -(p_\beta - p_\alpha)V_\beta + \gamma A,
\end{equation}
where $p_\beta$ and $p_\alpha$ represent the pressures of the respective phases.
Rigorously speaking, $p_\beta$ is the pressure of the hypothetical cluster defined such as possessing the bulk property and filling inside the surface of tension.
Its derivation was given in the literatures \cite{buff1951,nishioka1977,nishioka1987,debenedetti1998,corti2011}.
The present authors have given a transparent explanation for the volume term $W_\mathit{vol} = -(p_\beta - p_\alpha)V_\beta$ through a grand potential formalism recently \cite{morisub}.
That is, one can readily understand the form of $W_\mathit{vol}$ on the basis of the fact that the reversible work of formation of the critical nucleus is the grand potential difference (recall that the grand potential $\Omega$ is equal to $-pV$).
Nishioka and Kusaka \cite{nishioka1992a} found out that in the case that the $\beta$ phase is incompressible, the volume term $W_\mathit{vol} = -(p_\beta - p_\alpha)V_\beta$ can be rewritten in the form $n\Delta\mu$.
That is, they integrated $(\partial\mu/\partial p)_T = v$, which is nothing other than Gibbs-Duhem relation for the isothermal case, for the $\beta$ phase (the same procedure was followed by Debenedetti and Reiss \cite{debenedetti1998}).
Unfortunately, they concluded incorrectly that the form of Eq.~(\ref{eq:common}) was valid only for the case of the incompressible $\beta$ phase such as a nucleation of an incompressible liquid phase in a vapor phase.
One of the present authors has integrated $(\partial\mu/\partial p)_T = v$ for the $\alpha$ phase in case that the $\alpha$ phase is incompressible to get the form of Eq.~(\ref{eq:common}) \cite{morisub2}.
For example, this condition is valid for a bubble nucleation in an incompressible liquid phase.
For the case of the incompressible $\beta$ phase, we have
\begin{eqnarray}
\nonumber
W_\mathit{vol} &=& \frac{V_\beta}{v_\beta} [\mu_\beta(T,p_\alpha) - \mu_\beta(T,p_\beta)] \\
\label{eq:wvolbeta}
&=& \frac{V_\beta}{v_\beta} [\mu_\beta(T,p_\alpha) - \mu_\alpha(T,p_\alpha)],
\end{eqnarray}
where $T$ is the temperature, which is assumed to be uniform throughout the system.
To reach to last expression, we have used the fact that the chemical potential of the nucleating phase $\mu_\beta(T,p_\beta)$ is equal to that of the parent phase $\mu_\alpha(T,p_\alpha)$ (that is, the chemical potential is uniform throughout the system).
For the case of the incompressible $\alpha$ phase, we have
\begin{eqnarray}
\nonumber
W_\mathit{vol} &=& \frac{V_\beta}{v_\alpha} [\mu_\alpha(T,p_\alpha) - \mu_\alpha(T,p_\beta)] \\
\label{eq:wvolalpha}
&=& \frac{V_\beta}{v_\alpha} [\mu_\beta(T,p_\beta) - \mu_\alpha(T,p_\beta)].
\end{eqnarray}

For extension to the munticoponent system, we may merely follow Nishioka and Kusaka \cite{nishioka1992a}.
In this paper, following Nishioka and Kusaka \cite{nishioka1992a}, we will expand the chemical potentials for the case that no volume change is associated with the $\alpha$-$\beta$ phase transition.
We limit ourselves to the single component system for simplicity.

In Eqs.~(\ref{eq:wvolbeta}) and (\ref{eq:wvolalpha}), $\mu_\beta(T,p_\alpha) - \mu_\alpha(T,p_\alpha)$ and $\mu_\beta(T,p_\beta) - \mu_\alpha(T,p_\beta)$ are not the measurable quantity experimentally directly --- at least, the equation of state must be measured to be integrated.
The ^^ ^^ undercooling" (driving force) is defined as the difference between targeted chemical potential and that at the $\alpha$-$\beta$ equilibrium.
However, this terminology is somewhat confusing, because a temperature decrement is imagined thereby.
Let us use, instead, the chemical potential difference in this paper.
For a critical nucleus, as mentioned above, $\mu_\beta(T,p_\beta)$ and $\mu_\alpha(T,p_\alpha)$ is equal with each other and this value is the chemical potential of the reservoir.
For latter convenience, let us define the chemical potential of the reservoir $\mu_\mathit{re}$, which is the chemical potential of the real system.
Also we define the chemical potential at the $\alpha$-$\beta$ phase equilibrium, $\mu_\mathit{eq}$.
We define the chemical potential difference
\begin{equation}
\label{eq:deltamu}
\Delta\mu = \mu_\mathit{re} - \mu_\mathit{eq}.
\end{equation}
In the case that the $\alpha$ phase is a rarefied gas, for example, this quantity is expressed as
\begin{equation}
\label{eq:lnS}
\Delta\mu = k_BT\ln S.
\end{equation}
where $S \equiv p_\alpha/p_\mathit{eq}$ with $p_\mathit{eq}$ being the pressure at the $\alpha$-$\beta$ phase equilibrium.
Some ones inappropriately state that $\mu_\beta(T,p_\alpha) - \mu_\alpha(T,p_\alpha)$ is measurable; this is not entirely correct --- this quantity can be expressed as $\mu_\beta(T,p_\alpha) - \mu_\alpha(T,p_\alpha) = [\mu_\beta(T,p_\alpha) - \mu_\mathit{eq}] - [\mu_\alpha(T,p_\alpha) \ - \mu_\mathit{eq}] = v_\beta (p_\alpha-p_\mathit{eq}) - k_BT\ln(p_\alpha/p_\mathit{eq})$ for the case that the $\alpha$ phase is an ideal gas and the $\beta$ phase is incompressible (this is, however, not the present concern).
[We note that in the crystal nucleation form the melt $\Delta\mu = \Delta S (T_m-T)$ replaces ($\Delta S$ is the entropy of melting and $T_m$ the melting temperature).]
Correspondingly, misunderstandingly $\mu_\beta - \mu_\alpha$ in Eq.~(\ref{eq:common}) is substituted by Eq.~(\ref{eq:lnS}) and sometimes \cite{kelton,mcgraw1981,wilemski1991,yasuoka1998,holten2005,noguera2006,wasai2007,uwaha2010}
\begin{equation}
W^* = \frac{16\pi v^2 \gamma^3}{3\Delta\mu^2}
\end{equation}
is used for the barrier height of nucleation, which is obtained by solving $\partial W/\partial R =0$ for $R$ and inserting the solution into $W$.
This situation indicates that expressing $W$ in terms of the true driving force, $\Delta\mu$, is strongly desired.

The other subject of Nishioka and Kusaka \cite{nishioka1992a} was formulation of the reversible work for non-critical clusters.
Nishioka and Mori \cite{Nishioka1992b} rewrote their formula and gave an approximate concise expression.
A little later, Debenedetti and Reiss~\cite{debenedetti1998} drove a similar expression to that in Ref. \cite{nishioka1992a}.
This subject is, however, not the present concern. 
We note that for the system including a non-critical cluster, the chemical potential is no longer uniform.

\section{Calculations}
\label{sec:calc}

\begin{figure*}[p]
\includegraphics[width=12cm]{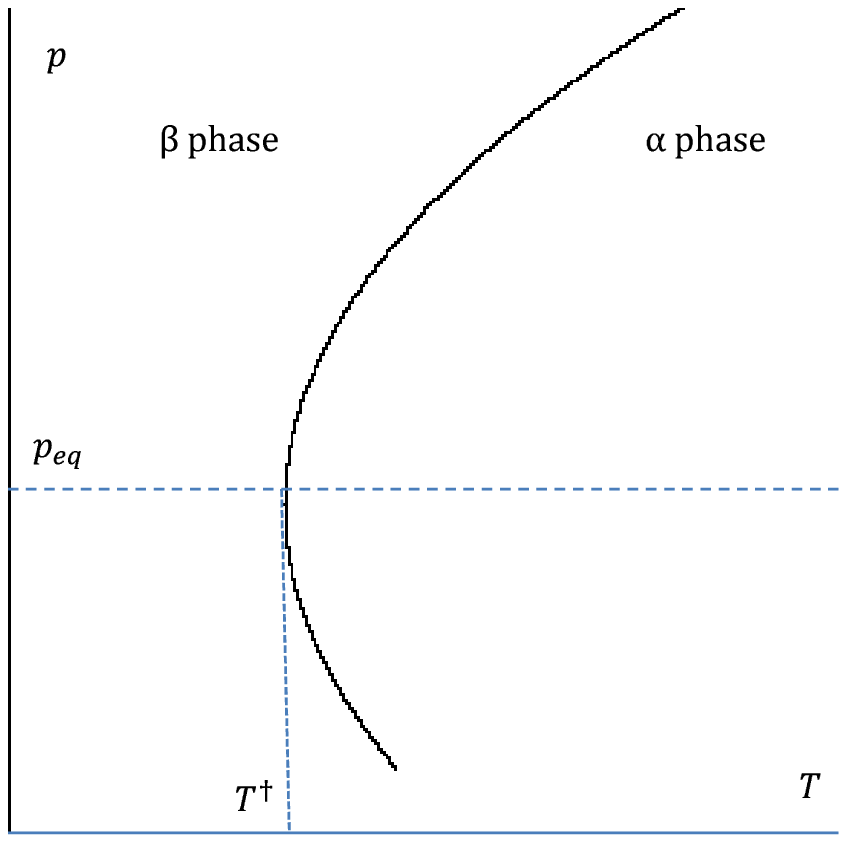}
\caption{\label{fig:pt} An illustration of the $p$-$T$ phase diagram near the point where no volume change is associated with the $\alpha$-$\beta$ phase transition.}
\end{figure*}

In the above way, one has known under what cases the form of Eq.~(\ref{eq:common}) is valid; those are not the cases which one intuitively imagines.
The purpose of this paper is to give a form of the volume term of $W$ in this intuitive case (the case that the $\alpha$-$\beta$ phase transition accompanies no volume change).
Let us start with the $p$-$T$ phase diagram.
The normal case is that the molecular volume of the low-temperature phase ($\beta$ phase) is smaller than that of the high-temperature phase ($\alpha$ phase).
In this case, the phase boundary is a curve with positive slope in the $p$-$T$ phase diagram.
The case that the molecular volume of the $\beta$ phase is larger than that of the $\alpha$ phase such as the water-ice case is an abnormal case.
In this case, the phase boundary has negative slope.
What we will consider is near the point connecting those two resumes.
For the case that the molecular volumes of two phases are the same over a certain region, because the phase boundary becomes a straight vertical line in this region in the $p$-$T$ phase diagram, the pressure no longer induces the phase transition.
For the case of water-ice, for example, while near an atmospheric pressure the water is denser than the ice, at hight pressure the ice becomes denser than the water.
The phase diagram is as illustrated in Fig.~\ref{fig:pt} (strictly speaking, this is a mere intuition --- the phase boundary in $p$-$T$ phase  diagram bends due to transition between the normal ice and a high-pressure one as will be discussed in detail in a latter part of Sec.~\ref{sec:discussion}).
Hereafter, we restrict ourselves on the temperature $T=T^\dag$ (we sometimes omit $T^\dag$ itself for brevity).
We define $p_\mathit{eq}$ as the point at which no volume change is associated with the $\alpha$-$\beta$ phase transition.
While in the region $p > p_\mathit{eq}$, the phase boundary behaves normally, in the region $p < p_\mathit{eq}$ it behaves abnormally.
In the both regions, the $\beta$ phase is more stable than the $\alpha$ phase.
This means that the pressure-induced phase transition from the $\alpha$ phase into the $\beta$ phase takes place.

\begin{figure*}[p]
\includegraphics[width=12cm]{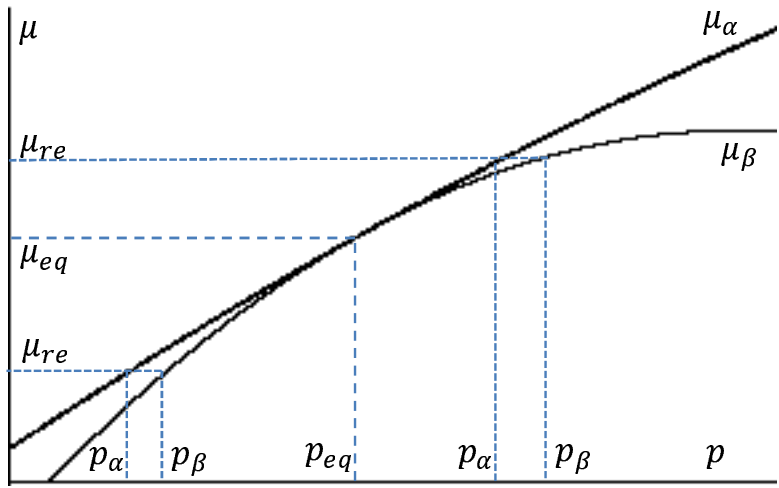}
\caption{\label{fig:mup} $\mu$-$p$ diagram around $p=p_\mathit{eq}$ at $T=T^\dag$.
$\mu$-$p$ relations for both $p_\alpha > p_\mathit{eq}$ and $p_\alpha < p_\mathit{eq}$ are schematically illustrated.}
\end{figure*}

$\mu(T^\dag,p)$ around $p=p_\mathit{eq}$ is schematically drawn in Fig.~\ref{fig:mup}.
Except for just on the point $p=p_\mathit{eq}$, the $\alpha$ phase is always metastable phase as mentioned above.
For thermodynamic consistency the $\alpha$ phase is less compressible than the $\beta$ phase ($\kappa_\alpha < \kappa_\beta$).
If one imagines a vapor-liquid case as a typical case, this observation is, at a glance, surprising because the valor phase is overwhelmingly compressible.
Considering, for example, a crystalline nucleation from a (well compressed) undercooled liquid or a glass phase, this situation may be reasonable; focusing on the free volumes of two phases, the crystalline solid phase may be more compressible.
For the vapor-liquid case, the situation illustrated in Fig.~\ref{fig:pt} is impossible.
Let us express the chemical potential in a series expansion around $p^\mathit{eq}$ in the pressure up to second order:
\begin{equation}
\label{eq:museries}
\mu(p)-\mu_\mathit{eq} = v_\mathit{eq} \left[(p-p_\mathit{eq}) -\frac{1}{2} \kappa (p-p_\mathit{eq})^2\right],
\end{equation}
where $v_\mathit{eq}$ is the molecular volume at the $\alpha$-$\beta$ phase transition and $\kappa$ the isothermal compressibility $-(1/v)(\partial v/\partial p)_T$.
This equation can be obtained by integrating $\kappa = -(1/v)(\partial v/\partial p)_T$ twice with help of $v = (\partial\mu/\partial p)_T$ as if $\kappa$ is constant and then expanding.
We obtain from Eq.~(\ref{eq:museries}) the series expansion solution to $p - p_\mathit{eq}$ as 
\begin{equation}
\label{eq:pseries}
p - p_\mathit{eq} = \frac{1}{v_\mathit{eq}} \left\{ \left[\mu(p)-\mu(p_\mathit{eq})\right] - \frac{\kappa}{2v_\mathit{eq}} \left[\mu(p)-\mu(p_\mathit{eq})\right]^2 \right\}.
\end{equation}
We note here that because Eq.~(\ref{eq:museries}) is a quadratic equation, we can solve it using the quadratic formula and then expand the solution assuming the smallness of $\kappa$ to obtain Eq.~(\ref{eq:pseries}) [one can have a solution $x = -c/b -ac^2/b^3 + O(a^2)$ to $ax^2 + bx + c = 0$ by expanding $x = (-b+\sqrt{b^2-4ac})/2a$ into a power series in $a$.].
Also, we can get Eq.~(\ref{eq:pseries}) by solving the quadratic equation iteratively assuming the smallness of the coefficient of the second order term [one can rewrite the quadratic equation $ax^2 + bx + c = 0$ into $x = -c/b - (a/b)x^2$ and then the 0th order solution $x^{(0)} = -c/b$ and the 1st order correction $x^{(1)} = -(a/b)(x^{(0)})^2 = -ac^2/b^3$].
Applying Eq.~(\ref{eq:pseries}) with $\kappa_\alpha$ and $\kappa_\beta$ begin the isothermal compressibilities at $p=p_\mathit{eq}$ for both phases, we have
\begin{eqnarray}
\label{eq:pbeta}
p_\beta - p_\mathit{eq} &=& \frac{1}{v_\mathit{eq}} \left[ (\mu_\mathit{re}-\mu_\mathit{eq}) + \frac{\kappa_\beta}{2v_\mathit{eq}} (\mu_\mathit{re}-\mu_\mathit{eq})^2 \right], \\
\label{eq:palpha}
p_\alpha - p_\mathit{eq} &=& \frac{1}{v_\mathit{eq}} \left[ (\mu_\mathit{re}-\mu_\mathit{eq}) + \frac{\kappa_\alpha}{2v_\mathit{eq}} (\mu_\mathit{re}-\mu_\mathit{eq})^2 \right].
\end{eqnarray}
Subtracting Eq.~(\ref{eq:palpha}) from Eq.~(\ref{eq:pbeta}), we have
\begin{equation}
\label{eq:pdiff}
p_\beta - p_\alpha = \frac{\kappa_\beta - \kappa_\alpha}{2v_\mathit{eq}^2} (\mu_\mathit{re}-\mu_\mathit{eq})^2.
\end{equation}
Here, we note that if we neglect the compressibility effect, i.e., in the first order in $\mu$, we have a result that no volume change, no pressure-induced phase transition.
Putting Eq.~(\ref{eq:pdiff}) into Eq.~(\ref{eq:exact}), the reversible work becomes
\begin{equation}
\label{eq:correct}
W = -\frac{\kappa_\beta - \kappa_\alpha}{2v_\mathit{eq}^2} (\mu_\mathit{re} - \mu_\mathit{eq})^2 V_\beta + \gamma A.
\end{equation}
This equation is, in appearance, of a form far from the commonly used formula [Eq.~(\ref{eq:common})]. 

In this way, we have shown that for the case of no volume change at the phase transition, for which the commonly used formula of Eq.~(\ref{eq:common}) seems intuitively to hold, the reversible work of formation of the critical nucleus takes a form far from Eq.~(\ref{eq:common}).
The volume term of the reversible work of formation of the critical nucleus possesses a form including a second order in $\Delta\mu \equiv \mu_\mathit{re} -\mu_\mathit{eq}$, not a first order.
The reason why the second order term appears is because the second order term is the lowest order term necessary for the self-consistent treatment.

\section{Discussions}
\label{sec:discussion}

Let us discuss a relation of form of Eq.~(\ref{eq:wvolbeta}) and the present results [Eq.~(\ref{eq:correct})].
Assuming that the deviation from the equilibrium is small, one can make the following expansion.
\begin{eqnarray}
\nonumber
p_\beta - p_\alpha &=& (p_\beta - p_\mathit{eq}) - (p_\alpha - p_\mathit{eq}) \\
\nonumber
&=& \left.\left(\frac{\partial p_\beta}{\partial\mu_\beta}\right)_T\right|_{\mu_\beta(p_\beta) = \mu_\mathit{eq}} [\mu_\beta(p_\beta)-\mu_\beta(p_\mathit{eq})] \\
\nonumber
&& -\left.\left(\frac{\partial p_\beta}{\partial\mu_\beta}\right)_T\right|_{\mu_\beta(p_\alpha) = \mu_\mathit{eq}} [\mu_\beta(p_\alpha)-\mu_\beta(p_\mathit{eq})] + \mbox{h.o.} \\
\nonumber
&=& \frac{1}{v_\mathit{eq}} [\mu_\beta(p_\beta) - \mu_\beta(p_\alpha)] + \mbox{h.o.} \\
&=& \frac{1}{v_\mathit{eq}} [\mu_\alpha(p_\alpha) - \mu_\beta(p_\alpha)] + \mbox{h.o.},
\end{eqnarray}
with $\mbox{h.o.}$ standing for higher order terms.
Accordingly, instead of Eq.~(\ref{eq:wvolbeta}), we have
\begin{equation}
\label{eq:wvoleq}
W_\mathit{vol} = \frac{V_\beta}{v_\mathit{eq}} [\mu_\beta(T,p_\alpha) - \mu_\alpha(T,p_\alpha)] + \mbox{h.o.}
\end{equation}
Applying Eq.~(\ref{eq:pseries}) we can also expand $\mu_\beta(T,p_\alpha) - \mu_\alpha(T,p_\alpha)$ as
\begin{equation}
\mu_\beta(T,p_\alpha) - \mu_\alpha(T,p_\alpha) = \frac{v_\mathit{eq}}{2} (\kappa_\beta - \kappa_\alpha) (p_\alpha-p_\mathit{eq})^2.
\end{equation}
Using this equation and Eq.~(\ref{eq:palpha}) we rewrite Eq.~(\ref{eq:wvoleq}) into Eq.~(\ref{eq:correct}).
In this way, we have an additional proof.
In addition, versatility of the form of Eq.~(\ref{eq:wvolbeta}) has been shown.

Here, let us discuss about the reality of the phase transition without volume change.
In Sec.~\ref{sec:calc} we have mentioned about the water-ice case.
In an early phase diagram in $p$-$T$ space, the phase boundary between the ice III and water has been drawn as vertical (see, \textit{e.g.} Fig.~5.15 of \cite{atkins}).
In updated ones such as in \cite{salzmann2006,salzmann2009,dunaeve2010,salzmann2011,fortes2012} that boundary is not vertical.
Possibilities of no volume change, however, remain for ice VII-ice VIII boundary such as drawn in the phase diagram \cite{salzmann2009,salzmann2011,fortes2012} and for ice III-ice IX transition regarding bond ordering \cite{knight2006}.
There are other possibilities regarding ice-ice phase boundaries which are less revealed.
A problem about those possibilities is that the both phases are solid states so that the phase transitions are slow.
One may, however, not be disappointed.
There exists certainly a point where no volume change is associated with the phase transition on the melting curve of graphite \cite{bundy1980,vanthiel1989,vanthiel1992,bundy1996}.
A note is that the slope of the melting curve is positive at low pressure and then becomes negative as the pressure increases.
Therefore, the shape of the melting curve is convex to the right as opposed to Fig.~\ref{fig:pt}.
The fundamental framework of the theory is, however, applicable.

\section{Concluding remarks}
\label{sec:conc}

This paper has made a new finding; a note should be added to some textbooks thereby.
Intuitively one regards that if the volume change at the phase transition can be neglected, the volume term of the reversible work of forming the critical nucleus becomes of the form $n(\mu_\beta - \mu_\alpha)$.
This intuition should be amended.
In this case, the volume term is of the second order in $\mu_\mathit{re} - \mu_\mathit{eq}$, not of the first order.
After Nishioka and Kusaka \cite{nishioka1992a}, little of concerning description of textbook has been rewritten.
The authors hope that repeating of the same situation regarding the present finding should be avoided.
Concretely speaking, the exact form of $\Delta\mu$ should be clearly written.
Without doing so, the misunderstanding that the form of Eq.~(\ref{eq:common}) would hold for the case of a phase transition without volume change may happen.

The present result may affect the application of the nucleation theorem \cite{kashchiev1982,oxtoby1994,bowles2001} because the original form is based on the form of $n\Delta\mu$.
Although  the $\Delta\mu$ is exactly given by  $\mu_\beta(T,p_\alpha) - \mu_\alpha(T,p_\alpha)$ and the volume term is given by Eq.~(\ref{eq:wvoleq}), practically $\partial \ln J / \partial \ln S$ is calculated such as in literatures \cite{holten2005,strey1994,ford1997,wolk2001,lummen2005}.
Therefore, reconsideration of the application of the nucleation theorem is necessary.
This is one of the future researches.

In this paper, we have treated the special case that the phase transition accompanies no volume change.
Detecting an anomalous behavior in reality is one of the future researches.
Consideration on the expression in terms of $\mu_\mathit{re} - \mu_\mathit{eq}$ for general cases is in progress.



%% References


\bibliographystyle{model1-num-names}
\bibliography{novchange}






% \end{thebibliography}

\newpage

\newpage

\end{document}